\documentclass[12pt,preprint,longabstract]{aastex}

\usepackage{verbatim}
\usepackage{subfigure}
\usepackage{epsfig}
\usepackage{graphicx}

\def\arcsec{\ifmmode '' \else $''$\fi}

\def\arcsecpoint{\ifmmode ''\!. \else $''\!.$\fi}

\def\kms{\ifmmode {\rm km\ s}^{-1} \else km s$^{-1}$\fi}
\def\Msun{\ifmmode {\rm M}_{\odot} \else M$_{\odot}$\fi}
\def\Lsun{\ifmmode {\rm L}_{\odot} \else L$_{\odot}$\fi}
\def\Zsun{\ifmmode {\rm Z}_{\odot} \else Z$_{\odot}$\fi}

\def\ergscm2{ergs\,s$^{-1}$\,cm$^{-2}$}
\def\icm3{{\rm cm}^{-3}}
\def\icm2{{\rm cm}^{-2}}
\def\qo{\ifmmode q_{\rm o} \else $q_{\rm o}$\fi}
\def\Ho{\ifmmode H_{\rm o} \else $H_{\rm o}$\fi}
\def\ho{\ifmmode h_{\rm o} \else $h_{\rm o}$\fi}
\def\ltsim{\raisebox{-.5ex}{$\;\stackrel{<}{\sim}\;$}}
\def\gtsim{\raisebox{-.5ex}{$\;\stackrel{>}{\sim}\;$}}
\def\vFWHM{\ifmmode v_{\mbox{\tiny FWHM}} \else
            $v_{\mbox{\tiny FWHM}}$\fi}
\def\CCF{\ifmmode F_{\it CCF} \else $F_{\it CCF}$\fi}
\def\ACF{\ifmmode F_{\it ACF} \else $F_{\it ACF}$\fi}
\def\Halpha{\ifmmode {\rm H}\alpha \else H$\alpha$\fi}
\def\Hbeta{\ifmmode {\rm H}\beta \else H$\beta$\fi}
\def\Hgamma{\ifmmode {\rm H}\gamma \else H$\gamma$\fi}
\def\Hdelta{\ifmmode {\rm H}\delta \else H$\delta$\fi}
\def\Lya{\ifmmode {\rm Ly}\alpha \else Ly$\alpha$\fi}
\def\Lyb{\ifmmode {\rm Ly}\beta \else Ly$\beta$\fi}
\def\Lyg{\ifmmode {\rm Ly}\beta \else Ly$\gamma$\fi}

\def\hi{H\,{\sc i}}

\def\heiii{He\,{\sc iii}}

\def\cii{C\,{\sc ii}}
\def\ciii{\ifmmode {\rm C}\,{\sc iii} \else C\,{\sc iii}\fi}
\def\civ{\ifmmode {\rm C}\,{\sc iv} \else C\,{\sc iv}\fi}
\def\cv{\ifmmode {\rm C}\,{\sc v} \else C\,{\sc v}\fi}
\def\cvi{\ifmmode {\rm C}\,{\sc vi} \else C\,{\sc vi}\fi}

\def\nv{N\,{\sc v}}

\def\o5007{[O\,{\sc iii}]\,$\lambda5007$}

\def\ovi{O\,{\sc vi}}

\def\mgii{Mg\,{\sc ii}}

\def\siiv{Si\,{\sc iv}}
\def\siIII{Si\,{\sc iii}}
\def\siII{Si\,{\sc ii}}

\def\siv{S\,{\sc iv}}

\def\nai{Na\,{\sc i}}

\def\o{\o}

\newcommand{\vy}[2]{#1_{\scriptscriptstyle #2}}
\def\gtorder{\mathrel{\raise.3ex\hbox{$>$}\mkern-14mu
             \lower0.6ex\hbox{$\sim$}}}
\def\ltorder{\mathrel{\raise.3ex\hbox{$<$}\mkern-14mu
             \lower0.6ex\hbox{$\sim$}}}
\def\proptwid{\mathrel{\raise.3ex\hbox{$\propto$}\mkern-14mu
             \lower0.6ex\hbox{$\sim$}}}

\begin{document}

\shortauthors{Edmonds et al.}
\shorttitle{Galactic Scale Absorption Outflow in IRAS F04250-5718}

\title{Galactic Scale Absorption Outflow in the Low Luminosity Quasar IRAS~F04250-5718: HST/COS Observations}

\author{
Doug~Edmonds\altaffilmark{1}\email{edmonds@vt.edu}, Benoit~Borguet\altaffilmark{1}, Nahum~Arav\altaffilmark{1}, Jay~P.~Dunn\altaffilmark{2}, Steve~Penton\altaffilmark{3}, Gerard~A.~Kriss\altaffilmark{4,5}, Kirk~Korista\altaffilmark{6}, Elisa~Costantini\altaffilmark{7}, Katrien~Steenbrugge\altaffilmark{8,9}, J.~Ignacio~Gonzalez-Serrano\altaffilmark{10}, Kentaro~Aoki\altaffilmark{11}, Manuel~Bautista\altaffilmark{6}, Ehud~Behar\altaffilmark{12}, Chris~Benn\altaffilmark{10}, D.~Micheal~Crenshaw\altaffilmark{13}, John~Everett\altaffilmark{14}, Jack~Gabel\altaffilmark{15}, Jelle~Kaastra\altaffilmark{7}, Maxwell~Moe\altaffilmark{16}, Jennifer~Scott\altaffilmark{17}
}

\altaffiltext{1}{Department of Physics, Virginia Tech, Blacksburg, VA 24061, USA}
\altaffiltext{2}{Department of Chemistry and Physics, Augusta State University, Augusta, Georgia 30904-2200, USA}
\altaffiltext{3}{Center for Astrophysics and Space Astronomy, University of Colorado, Boulder, Colorado 80309-0389, USA}
\altaffiltext{4}{Space Telescope Science Institute, 3700 San Martin Drive, Baltimore, MD 21218, USA}
\altaffiltext{5}{Department of Physics and Astronomy, The Johns Hopkins University, Baltimore, MD 21218, USA}
\altaffiltext{6}{Department of Physics, Western Michigan University, Kalamazoo, MI 49008-5252, USA} 
\altaffiltext{7}{SRON, Netherlands Institute for Space Research, Sorbonnelaan 2, 3584CA Utrecht, The Netherlands}
\altaffiltext{8}{Instituto de Astronom\'ia, Universidad Cat\'olica del Norte, Avenida Angamos 0610, Casilla 1280, Antofagasta, Chile}
\altaffiltext{9}{Department of Physics, University of Oxford, Keble Road, Oxford OX1 3RH, UK}
\altaffiltext{10}{Instituto de Fisica de Cantabria (CSIC), 39005, Santander, Spain}
\altaffiltext{11}{Subaru Telescope, National Astronomical Observatory of Japan, 650 North A’ohoku Place, Hilo, HI 96720, USA}
\altaffiltext{12}{Department of Physics, Technion Israel Institute of Technology, Haifa 32000, Israel}
\altaffiltext{13}{Department of Physics \& Astronomy, Georgia State University, Atlanta, GA 30303, USA}
\altaffiltext{14}{Department of Astronomy, University of Wisconsin–Madison, 475 N. Charter Street, Madison, WI 53706-1582, USA}
\altaffiltext{15}{Physics Department, Creighton University, Omaha, NE 68178, USA}
\altaffiltext{16}{Department of Astronomy, Harvard University, Cambridge, MA 02138, USA}
\altaffiltext{17}{Department of Physics, Astronomy, and Geosciences, Towson University, Towson, MD 21252, USA}

\begin{abstract}


We present absorption line analysis of the outflow in the quasar IRAS~F04250-5718. Far-ultraviolet data from the Cosmic Origins Spectrograph onboard the Hubble Space Telescope reveal intrinsic narrow absorption lines from high ionization ions (e.g., \civ, \nv, and \ovi) as well as low ionization ions (e.g., \cii\ and \siIII). We identify three kinematic components with central velocities ranging from $\sim$~-50 to $\sim$~-230~km~s$^{-1}$. Velocity dependent, non-black saturation is evident from the line profiles of the high ionization ions. From the non-detection of absorption from a metastable level of \cii, we are able to determine that the electron number density in the main component of the outflow is $\ltsim$ 30 cm$^{-3}$. Photoionization analysis yields an ionization parameter log~$U_H \sim -1.6 \pm 0.2$, which accounts for changes in the metallicity of the outflow and the shape of the incident spectrum. We also consider solutions with two ionization parameters. If the ionization structure of the outflow is due to photoionization by the active galactic nucleus, we determine that the distance to this component from the central source is $\gtsim$ 3 kpc. Due to the large distance determined for the main kinematic component, we discuss the possibility that this outflow is part of a galactic wind.
\end{abstract}


\section{Introduction}

Present in the UV spectra of $\sim$50\% of Seyfert I galaxies are strong absorption troughs \citep{Dunn07} with full-width at half-maximum $\sim$~20 to 400~km~s$^{-1}$ \citep{Crenshaw03} that are blue-shifted with respect to their emission line counterparts indicating material that is outflowing from the central source. These absorption spectra show lines of \civ~$\lambda \lambda$1548,1551; \nv~$\lambda \lambda$1239,1243; \ovi~$\lambda \lambda$1032,1038; and \hi~$\lambda$1216 with lower ionization species such as \siiv~$\lambda \lambda$1394,1403; and \mgii~$\lambda \lambda$2796,2804 seen less often \citep{Crenshaw99}. Narrow absorption lines (NALs), defined as having FWHM~$<$~500~km~s$^{-1}$, are also found in quasar spectra. For example, \citet{Misawa07} find intrinsic NALs in spectra of $\sim$50\% of their sample of optically selected quasars at redshifts $z=2-4$. Most studies of NAL quasars have focused on very luminous objects (e.g., \citealt{Ganguly99,Ganguly01,Hamann01,Hamann11,Misawa07,Misawa08}). NALs provide important diagnostics of outflows in active galactic nuclei (AGNs). For example, we can learn about chemical abundances in outflows (e.g., \citealt{Hamann97a,Gabel06,Arav07}) and provide estimates of their mass flow rates and kinetic luminosities, which are relevent to AGN feedback models (e.g., \citealt{deKool01,Hamann01,Moe09,Dunn10a,Dunn10b,Arav11}).

Most likely, the ionization structure of the outflowing gas is due to photoionization rather than collisional ionization (e.g., \citealt{Hamann99}). Therefore, photoionization studies are integral to understanding the outflow phenomenon. Computer codes such as {\sc Cloudy} \citep{Ferland98} have been developed to self-consistently solve the ionization and thermal equilibrium equations \citep{AGN^3}. Two parameters of particular interest are the total hydrogen column density ($N_H$) of the absorbing material and the ionization parameter ($U_H$), which is proportional to $\left( R^2 \vy{n}{H} \right) ^{-1}$ where $R$ is the distance to the absorber from the central source, and $\vy{n}{H}$ is the total hydrogen number density. Thus, determining $U_H$ and measuring $\vy{n}{H}$ yield the distance of the outflow. Previous studies have found outflows from AGN on scales of $\sim$~0.1~pc to tens of kpc from the central source (e.g., \citealt{Arav99b,Hamann01,Kraemer01,Moe09,Dunn10a}).

IRAS~F04250-5718 is the first of six objects we observed with the Cosmic Origins Spectrograph (COS, \citealt{Osterman10}) onboard the Hubble Space Telescope (HST), as part of our program aiming at determining the cosmological impact of AGN outflows (PID: 11686, PI: Arav). IRAS~F04250-5718 was chosen for observation because it shows little or no blending of absorption features, is bright enough to acquire high S/N in several orbits with the high throughput of COS, and has a redshift ($z=0.104$; \citealt{Thomas98}) that allows us to cover Ly$\alpha$ and Ly$\beta$ as well as \civ, \nv, and \ovi\ doublets with the COS far-ultraviolet (FUV) gratings. Observation of Ly$\alpha$ and Ly$\beta$ is necessary for absolute abundance determinations \citep{Arav07}, which will be presented in a future paper. Connected with our program, simultaneous XMM-Newton data were acquired (Costantini et al., in preparation).

IRAS~F04250-5718 is a radio-quiet, low luminosity quasar (with absolute blue magnitude $M_B \approx -24.7$; \citealt{Veron-Cetty06}). Due to the low luminosity and redshift ($z=0.104$),  IRAS~F04250-5718 is on the quasar/Seyfert borderline and is spectrally classified as a type 1.5 Seyfert \citep{Veron-Cetty06}. After some confusion due to erroneous positions recorded in some finding charts it was confirmed that the position of this galaxy is consistent with that of the faint blue star LB~1727 (see \citealt{Halpern98,Turner99}). Observational history: IRAS~F04250-5718 was discovered as an X-ray source (1H~0419-577) in the First HEAO survey \citep{Wood84} and, later, in the \textit{Einstein} Slew Survey \citep{Elvis92} and the \textit{Swift} BAT Survey \citep{Tueller08}. In a study identifying optical counterparts to HEAO X-ray sources, IRAS~F04250-5718 was identified with a Seyfert galaxy \citep{Brissenden87}. Observations using the Extreme Ultraviolet Explorer (EUVE) revealed that this object is extremely bright in the 50-180~\AA\ range \citep{Marshall95}. Most of the work on IRAS~F04250-5718 to date has focused on properties of the X-ray spectral energy distribution (e.g., \citealt{Guainazzi98,Turner99,Turner09,Pounds04a,Pounds04b}). Using Suzaku observations, \citet{Turner09} found evidence of a Compton-thick partial-covering absorber with inferred distances closer to the source than the Broad Line Region (BLR). Data in the far UV were obtained with the Far Ultraviolet Spectroscopic Explorer (FUSE) \citep{Dunn07,Wakker09}. Optical spectra were obtained using the Australian National University (ANU) telescope and the Anglo-Australian Telescope (AAT) as well as from the European Southern Observatory (ESO) (\citealt{Turner99} and references therein). To date, no detailed absorption line studies in the UV have been reported.


The plan of the paper is as follows: The observations are discussed in section~2. Column density determinations are presented in section~3. In section~4, we discuss photoionization analysis of the object, and in section~5, we derive the distance, mass, mass flow rate, and kinetic luminosity implied by our analysis. A discussion of our results is given in section~6 followed by a summary in section~7. The online version of this paper contains a complete spectrum of IRAS~F04250-5718 with absorption line identifications.


\section{Observations}

IRAS~F04250-5718 was observed for 8 orbits through the COS FUV channel using the Primary Science Aperature (PSA) and the medium resolution (R $\sim$ 18000) G130M and G160M gratings leading to a total integration time of $\sim 19000$~s and $\sim 15000$~s, respectively. Details about the COS on-orbit performances can be found in \citet{Osterman10}. Different central wavelength settings (3 for G130M and 4 for G160M) were used for each exposure in order to fill the inter-segment detector gap and minimize fixed-pattern noise, providing a continuous spectral coverage from roughly 1130 to 1795~\AA.

%

\begin{figure}[ht]
 \caption{Spectrum of IRAS04250-5718 obtained by COS in June 2010. Absorption troughs related to the intrinsic absorber are labeled. Here, $\sim$80\% of the spectrum is shown: A full identification of all the absorption features is presented in the online version.}
 \includegraphics[angle=90,width=0.9\textwidth]  {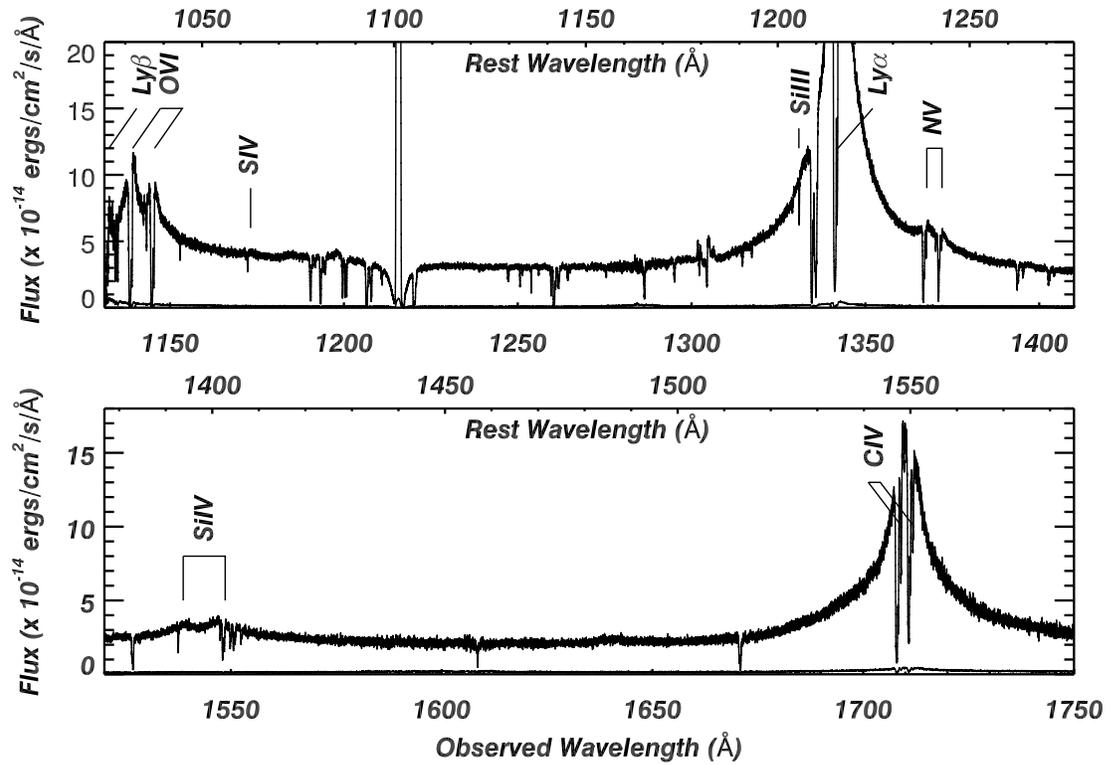}\\
 \label{fig:spec}
\end{figure}

The reduced data, processed by the COS calibration pipeline CALCOS\footnote{See details in HST Data Handbook for COS at http://www.stsci.edu/hst/cos/documents/handbooks /datahandbook/COS\_longdhbTOC.html }, were retrieved from the HST archive and combined together using IDL routines\footnote{The routines can be found at http://casa.colorado.edu/$\sim$danforth/costools.html along with an extensive description.} developed by the COS GTO team specifically for COS FUV data. These routines essentially perform flat-fielding, alignment, and co-addition of the processed exposures (see \citealt{Danforth10} for details). The wavelength calibration of each
combined segment is accurate to $<$~15~km~s$^{-1}$ \citep{Osterman10}. The final combined FUV spectrum of IRAS~F04250-5718 rebinned to a 20~km~s$^{-1}$ resolution grid has an average signal to noise ratio of $\gtsim 50 $. Figure~\ref{fig:spec} shows the majority of the COS spectrum (with the original $\sim$~2~km~s$^{-1}$ oversampling) along with identification of intrinsic absorption features. The online version of this paper contains a complete spectrum showing details of the line profiles along with identification of Galactic ISM absorption features.

The FUV spectrum of IRAS~F04250-5718 displays multiple absorption features from low ionization (e.g., \cii, \siIII) to high ionization (e.g., \civ, \ovi) ions. Due to the shallowness of the \cii\ trough (residual intensity $\gtsim$ 0.8), we inspected each data set individually and found that the absorption feature appears in each subexposure in regions seemingly free of flat-fielding issues. 
We identify three absorption components associated with the UV outflow in the un-rebinned COS spectrum of IRAS~F04250-5718. These three kinematic components, identified using the strong, unblended, \civ\ and \nv\ doublet lines, have their centroids located at velocities $v_1 \sim -38$~km~s$^{-1}$, $v_2 \sim -156$~km~s$^{-1}$, and $v_3 \sim -220$~km~s$^{-1}$ in the rest frame of the AGN ($z=0.104$).

\section{Data Analysis}

  \subsection{The unabsorbed emission model}
  \label{emimodele}

Determining the column densities associated with a given absorption trough requires knowledge of the unabsorbed AGN emission as well as which emission sources are covered by the outflow (e.g., \citealt{Gabel05}). The UV emission in AGNs mainly arises from three sources; the continuum, the broad emission line region (BELR), and the narrow emission line region (NELR). After correcting the spectrum for Galactic extinction using E(B-V)=0.0134 \citep{Schlegel98} and $R_V=3.1$ and shifting it to the rest frame, the continuum of IRAS~F04250-5718 is fitted with a single power law of the type $F_{\lambda} = F_{1100} (\lambda/1100)^{\beta}$ by performing a $\chi^{2}$ minimization over the regions free from known emission or absorption lines. A good fit of the continuum of IRAS~F04250-5718 was obtained using $F_{1100}=(3.613 \pm 0.004) \times 10^{-14}$ and $\beta=-1.603 \pm 0.091$.

In developing an emission model for the BELR and the NELR, we apply the restrictive approach outlined in \citet{Arav02} in which the emission lines are modeled using a minimum amount of free parameters. We are able to fit the main emission features (\civ, \ovi, and Ly$\alpha$) using two Gaussian components for the BELs characterized by a full-width at half-maximum (FWHM) of $\sim 3000$ and $\sim 9000$~km~s$^{-1}$, though a third, broader component (FWHM~$\sim 14000$~km~s$^{-1}$) was required to fit the blue wing of the Ly$\alpha$ emission. A spline fit was used for the \nv\ line given the difficulty in separating its shallow emission from the strong red wing of the Ly$\alpha$ BEL. For \ovi\ and \civ, we modeled the NEL of each line of the doublet by an additional Gaussian of FWHM~$\sim 800$~km~s$^{-1}$ (fixed by the width of the \ovi\ NEL), with the separation of the Gaussians fixed by the difference in wavelengths between the doublet lines. A single Gaussian of similar FWHM was used to model the Ly$\alpha$ NEL. The remaining weaker emission features in the spectrum were modeled by a spline fit.

\begin{figure}
  \includegraphics[angle=90,width=1.0\textwidth]  {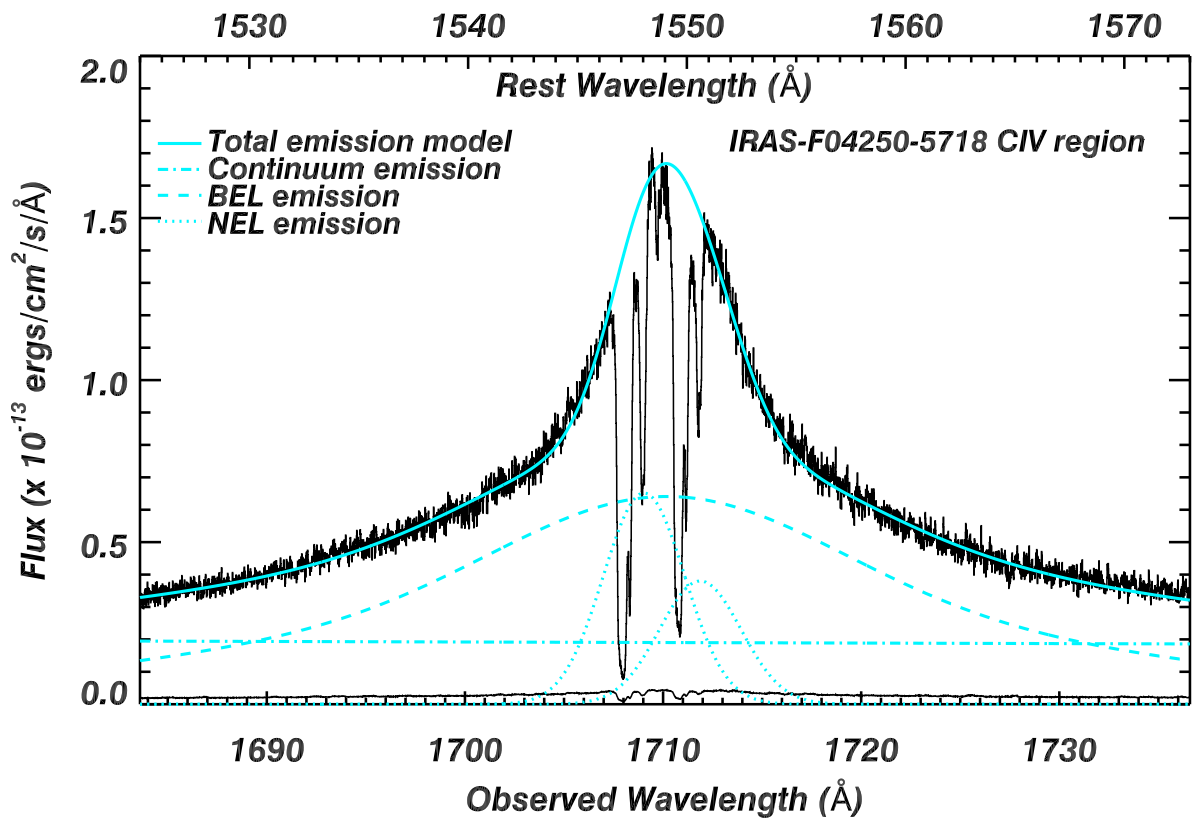}\\
 \caption{Details of the unabsorbed emission model for the \civ\ line in IRAS~F04250-5718. The total emission model is
  plotted as a solid line on top of the data. This model contains a contribution from three different sources;
  a continuum (dotted dashed line), a broad emission line (dashed line) and two narrow emission lines (dotted lines). The flux of the deepest absorption trough is lower than each emission component indicating that the absorption must cover all emission sources.}
 \label{em_compo_civ} 
\end{figure}




The results of the emission modeling process is presented in Figure~\ref{em_compo_civ} for \civ\ and online Figures~\ref{em_compo_civ}.1~-~\ref{em_compo_civ}.3 for the other strong emission features observed in the IRAS~F04250-5718 spectrum. Individual contributions of each emission source (continuum, BEL, and NEL) as well as the total model is plotted as a function of wavelength. In a case where the absorber covers only the continuum source and the BELR, we expect to find the absorption troughs above the NEL model (see \citealt{Arav02}). However, we observe that the flux of the deepest \civ\ absorption trough is lower than each emission component indicating that the absorption must cover all three emission sources.




\subsection{Column Density Determinations}
\label{sec:coldens}

   A first step in characterizing the physical properties of the absorber is the determinination of ionic
   column densities associated with the absorption troughs seen in the spectrum. 
   Early studies of intrinsic AGN absorption troughs using UV doublets such as \civ~$\lambda \lambda$1549,1551~\AA\ revealed that they rarely follow the expected 1:2 optical depth ratio between the weakest and strongest
   components of the doublet (e.g. \citealt{Barlow97b}). One interpretation of this
   observation is that the absorber does not totally cover the emission source (e.g., \citealt{Hamann97b,Ganguly99, Arav99b,Arav02,Gabel03,Arav05,Arav08}).
   Assuming a partial covering model of a single homogeneous emission source, the residual intensity observed
   for line $j$ of a given ion as a function of the radial velocity $v$ can be expressed as:
   \begin{equation}
     \label{eqcov}
     {I_j(v) = I_0(v) ((1-C_{ion}(v)) + C_{ion}(v) e^{-\tau_j(v)})}
   \end{equation}
   where $C_{ion}$ is the fraction of the emission source covered by an absorber having an optical depth $\tau_j$,
   and $I_0$ is the unabsorbed emission. While Equation~(\ref{eqcov}) is underconstrained for singlets (unless
   assumptions are made about $C_{ion}$), multiple lines of a given ion sharing the same lower energy level
   allow us to derive a velocity dependent solution for both $C_{ion}$ and $\tau_j$ by forcing
   their optical depth ratio $R=\tau_i/\tau_j$ to be identical to the one observed in the laboratory
   $R=\lambda_i f_i/\lambda_j f_j$. Variations of this technique, including
   the allowance of different covering fractions for each emission source, have been investigated in the literature \citep{Ganguly99,Gabel05}. However, the methods described in those papers cannot be applied to doublet lines unless one makes assumptions about the emission source/absorber model (generally assuming the same individual emission source covering for all ions
   in order to get enough constraints on the model).
   Note that not properly taking into account this partial covering usually results in
   an underestimation of the actual column densities (e.g. \citealt{Arav99b}), thus 
   hampering the interpretation of the physical properties of the intrinsic absorber.

   While the partial covering model easily accounts for the departure from an optical depth ratio of 1:2 between the lines of typical doublets, the physical validity of an absorber model displaying a step-function distribution of material accross the emission source is questionable. The fact that we typically find different covering fractions for different ions implies that the actual distribution of absorber material deviates from this simple absorber model. Alternative models in which the emission source is fully covered by a smooth distribution of absorbing material accross its spatial extension were studied in \citet{deKool02}, \citet{Arav05}, and \citet{Arav08}. Here we use the power law absorber model in which the distribution of optical depth accross
   the source is characterized by $\tau(x) = \tau_{max} x^{a}$. In this case, the normalized residual intensity observed for line $j$ of a given ion as function of the radial velocity $v$ can be expressed as
   \begin{equation}
     \label{eqpar}
     {I_j(v) = \int^1_0  e^{-\tau_{max,j}(v) x^{ a_{ion}(v)} } dx }
   \end{equation}
   Like Equation~(\ref{eqcov}), this equation is underconstrained for singlets (unless
   assumptions are made about the shape of the absorbing material distribution parametrized by $a_{ion}$)
   while a velocity dependent solution can be derived for both $a_{ion}$ and $\tau_{max,j}$ for
   multiple lines of a given ion sharing the same lower energy level by forcing
   their optical depth ratio $R=\tau_i/\tau_j$ to be identical to the one observed in the laboratory.

   Once an optical depth solution is obtained from either model, the velocity dependent column
   density is computed using the relation (adapted from \citealt{Savage91}):
   \begin{equation}
    \label{eqcoldens}
     {N_{ion}(v) = \frac{3.8 \times 10^{14}}{f_j \lambda_j}  \langle \tau_j(v) \rangle ~~(\textrm{cm}^{-2}\textrm{ km}^{-1}\textrm{ s})}
   \end{equation} 
   where $\lambda_j$ is the wavelength of the line (in \AA), $f_j$ is its oscillator strength, and
   $\langle \tau_j(v) \rangle$ is the spatially averaged optical depth across the emission source: For the partial covering model, $\langle \tau_j(v) \rangle = C_{ion}(v) \tau_j(v)$, and for the power law distribution, $\langle \tau_j(v) \rangle = \tau_{max,j}(v)/(a_{ion}(v) + 1)$.

%
%

   \subsubsection{Column densities of the metals}
\label{sec:metalcoldens}

\begin{figure}[ht]
  \centering
  \caption{Normalized absorption troughs detected in the IRAS~F04250-5718 spectrum.
   The strongest line of a doublet/multiplet is plotted in black while the weakest is plotted in gray (red in the online version). 
   For each doublet line as well as the \hi\ lines, we indicate the range of physically
   allowed residual intensity for the strongest line by plotting a
   dotted line showing the expected residual intensity for the strongest component if the emission source
   was fully covered by the absorber (optical depth ratio of  $R=\lambda_i f_i/\lambda_j f_j$).
   The Ly$\beta$ is only partially detected in our COS observations due to its location
   at the edge of the detector.}
  \includegraphics[angle=90,width=0.9\textwidth]  {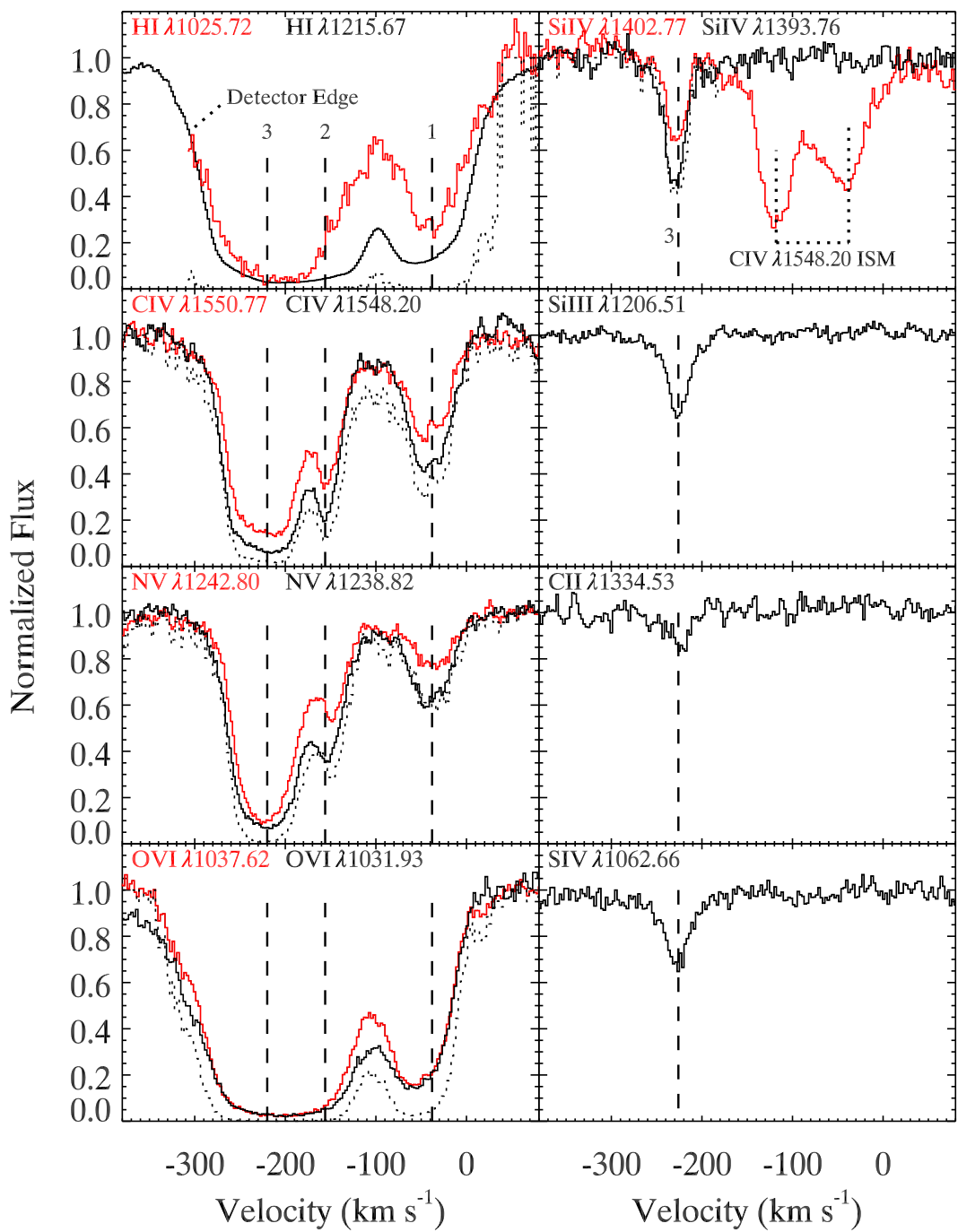}\\
  \label{figlineprof}
\end{figure}

In Figure~\ref{figlineprof}, we show the normalized absorption troughs detected in the IRAS~F04250-5718 spectrum. The line profiles are presented using the COS sampling of $\sim$~2~km~s$^{-1}$. In order to compute column densities across the trough, the line profiles are rebinned to a common velocity grid with resolution $\sim$~20~km~s$^{-1}$, slightly larger than the nominal resolution for COS. In determining integrated column densities, we combine kinematic components 2 and 3 since these components are seen only in \civ\ and \nv, but not in Ly$\alpha$, Ly$\beta$, or \ovi. Integrated column densities using the apparent optical depth (AOD) method, the partial covering (PC) method, and the Power law model (PL) for components 1 and 2+3 are given in Tables \ref{tabcold1} and \ref{tabcold2}, respectively. For doublets, the residual intensity equation (Equation~(\ref{eqcov}) for partial covering and Equation~(\ref{eqpar}) for the power law model) is solved for both lines simultaneously. For comparison, we derive column densities using the apparent optical depth method (AOD; $C_{ion}=1$) on the weaker line of the doublet. The velocity ranges used for integration of column densities are +30 to -90~km~s$^{-1}$ for component 1 and -90 to -290~km~s$^{-1}$ for component 2+3. The high velocity limit (-290~km~s$^{-1}$) is set by the partial observation of the Ly$\beta$ line profile in the COS spectrum due to its location at the edge of the detector. In $\S$\ref{sec:h1col}, we discuss our treatment of this line. For ions that only appear in component 3, we reduce the integration range to -210 to -290~km~s$^{-1}$ to avoid adding additional noise from regions where no trough is detected.

Absorption lines from \civ~$\lambda \lambda$1548,1551; \nv~$\lambda \lambda$1239,1243; and \ovi~$\lambda \lambda$1032,1038 are present in all three kinematic components. Since \civ\ and \nv\ do not show signs of strong saturation, both partial covering and power law methods yield reliable column density measurements for these ions for each respective absorber model. \ovi, on the other hand, shows signs of strong 
saturation not only in the core of each trough, but also in the wings, where a 1:1 ratio
is observed between the residual intensities of each line of the doublet. Strong saturation implies a large deviation of the actual column density in the absorber from AOD measurements.
The partial covering method also yields only a lower limit to the actual column density of \ovi\ due to the upper limit we impose on the optical depth of the weaker line: We limit the optical depth of the weaker line to $\tau \le 4$ because the difference between residual intensities of the lines is well within the noise whenever the optical depth of the weaker line of the doublet is $\gtsim 4$. Similar constraints are imposed on the power law method ($\langle \tau \rangle \le 4$ for the weaker line of the doublet) resulting in a column density that is only $\sim$20\% larger than that estimated by the partial covering method.

Absorption lines from \cii~$\lambda$1335; \siIII~$\lambda$1207; \siiv~$\lambda \lambda$1403,1394; and \siv~$\lambda$1063 are present only in kinematic component 3. For \siiv, we find that the residual intensity of the blue line of the doublet closely matches that of the red line when properly scaled by the ratio of line strengths ($I_B=I_R^2$), indicating that the AOD method yields reliable optical depths for this shallow trough. We then assume that this is also the case for the even shallower troughs from \cii, \siIII, and \siv.


\clearpage

   \subsubsection{Column density of \hi}
\label{sec:h1col}

The Ly$\beta$ trough is located near the edge of the blue detector ($\sim$1131~\AA) and only partially observed. Given the poor wavelength solution at these low wavelengths, we shifted the entire Ly$\beta$ trough by 39~km~s$^{-1}$ to match the kinematic structure of Ly$\alpha$. The poor wavelength solution may affect the shape of the line profile as well. In order to assess whether the Ly$\beta$ trough can be used in our analysis, we compared the COS data with 2006 FUSE observations of IRAS~F04250-5718. The FUSE spectra were downloaded from the Multimission Archive at Space Telescope (MAST) and processed with CalFUSE v3.2.3 \citep{Dixon07}. Due to low photon counts, we opted to use both the night and day exposures in the coadded spectrum since the dayglow emission features \citep{Feldman01} did not affect our scientific goals. We flux calibrated all eight segments to match the flux level in the LiF 1a segment. Comparison of the overlapping regions of the 2010 June COS and 2006 May FUSE spectra shows no variations in the continuum level or line profiles at the S/N of the FUSE data. Similarly, the shapes of the Ly$\beta$ troughs in the two instruments are identical within the noise. We conclude that a simple wavelength shift of the Ly$\beta$ trough in the COS data suffices for our analysis.

\hi\ column densities are derived by treating the Ly$\alpha$ and Ly$\beta$ lines as a doublet, and comparisons are made to AOD measurements using Ly$\beta$. The lines are clearly separated in component 1 allowing for reliable estimates of the \hi\ column density in this component for respective absorber models. 
In component 2+3, the deepest parts of the Ly$\alpha$ and Ly$\beta$ troughs look close to saturation, having a nearly  1:1 residual intensity ratio.
Inspection of the Lyman series lines in the FUSE spectrum indicates a possible saturation of the Ly$\beta$ and Ly$\gamma$ lines. Further, it is difficult to estimate the unabsorbed emission in the region of the Ly$\beta$ absorption since only part of the emission is observed and the COS S/N is severely degraded at the edge of the detector. We therefore consider the column density of \hi\ derived from the partial covering method a lower limit for kinematic component 2+3. The column density derived from the power law method is $\sim 50$\% higher. The lower S/N of the FUSE data does not allow us to get a better estimate of the \hi\ column density or determine a better unabsorbed emission model. However, the absence of a bound-free edge for \hi\ in the FUSE data allows us to place an upper limit on the \hi\ column density of $10^{15.8}$ cm$^{-2}$.

\begin{figure}[ht]
  \centering
  \caption{Comparison between the FUSE and HST/COS spectrum of IRAS-F04250-5718 taken four years apart. No noticeable changes in the continuum
  level or absorption line profiles is observed within the FUSE S/N between these two epochs though some variations may have occured
  between observations. The Ly$\beta$ trough observed in the FUSE data has a line profile that is identical
  within the noise to the one observed in the COS observation, suggesting that the COS wavelength solution in this region can be, as a first approximation, corrected
  by a simple shift of the line profile.}
  \includegraphics[angle=90,width=1.0\textwidth]  {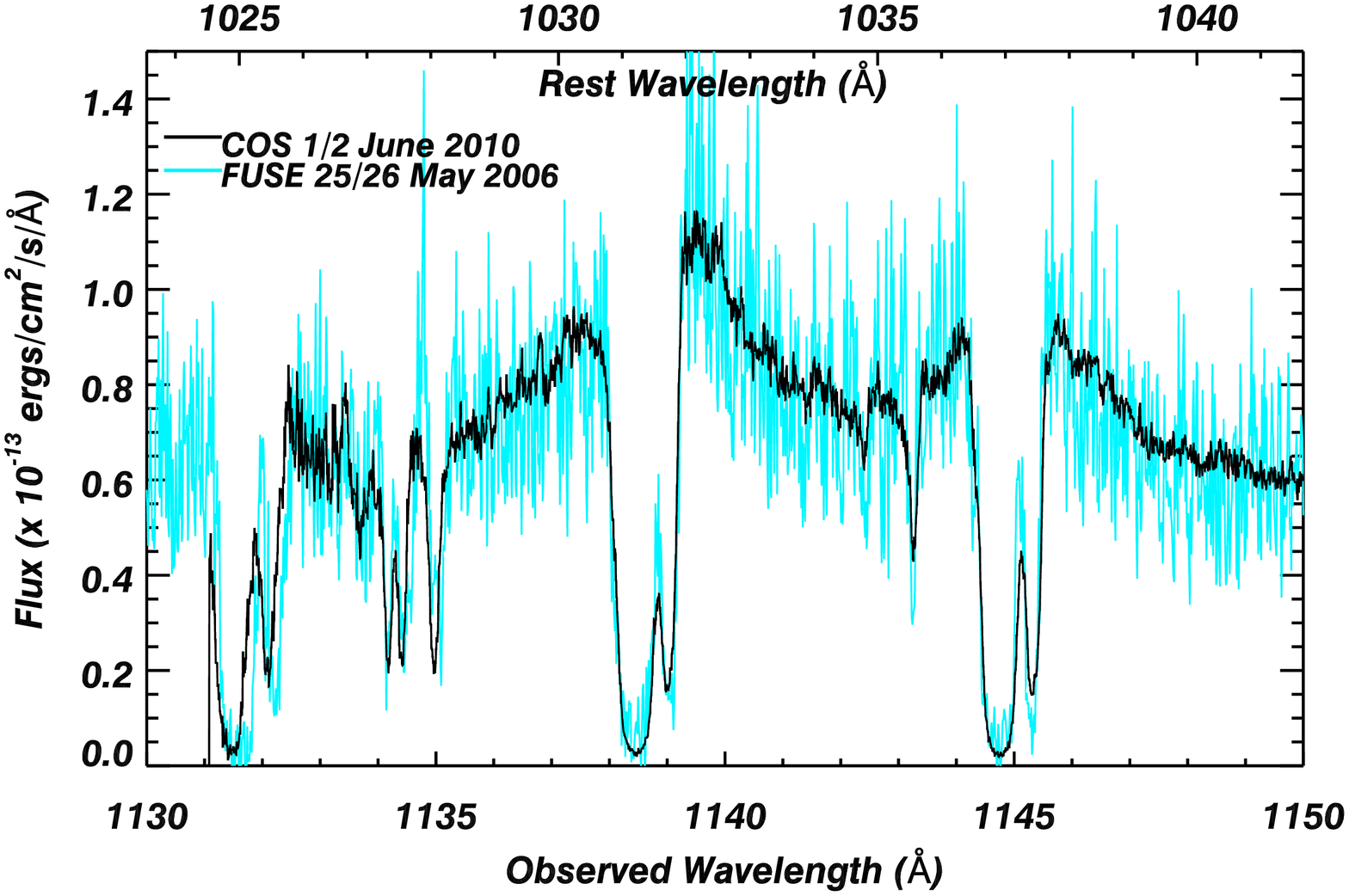}\\
  \label{compa_cosfuse}
\end{figure}

\begin{deluxetable}{lccc}
 \tablewidth{0.8\textwidth}
 \tablecolumns{4}
 \tabletypesize{\footnotesize}
 \tablecaption{Column densities for component 1 of IRAS~F04250-5718}
 \tablehead{
 \colhead{Ion} &
 \colhead{$N_{ion}$ (AOD)} &
 \colhead{$N_{ion}$ (PC)} &
 \colhead{$N_{ion}$ (PL)}
  }
 \startdata

\hi & 442$^{+16}_{-15}$ & 532$^{+25}_{-20}$ & 1340$^{+530}_{-5}$ \\
\civ & 79.5$^{+1.2}_{-1.2}$ & 113$^{+34}_{-11}$ & 304$^{+206}_{-2}$ \\
\nv &  59.7$^{+1.6}_{-1.6}$  & 73.2$^{+10.4}_{-3.7}$ & 135$^{+43}_{-26}$ \\
\ovi & 634$^{+5}_{-5}$  &   1220$^{+60}_{-30}$ & 2660$^{+1090}_{-0}$\\

 \enddata
\tablenotetext{~}{{\bf Notes.} Column densities are given as linear values in units of $10^{12}$ cm$^{-2}$. Asymmetrical errors are from photon statistics only and are derived using the technique outlined in \citet{Gabel05}.}
 
 \label{tabcold1}
\end{deluxetable}

\begin{deluxetable}{lccc}
 \tablewidth{0.8\textwidth}
 \tablecolumns{4}
 \tabletypesize{\footnotesize}
 \tablecaption{Column densities for component 2+3 of IRAS F04250-5718}
 \tablehead{
 \colhead{Ion} &
 \colhead{$N_{ion}$ (AOD)} &
 \colhead{$N_{ion}$ (PC)} &
 \colhead{$N_{ion}$ (PL)}
  }
 \startdata

\hi & 1730$^{+60}_{-50}$  &  2100$^{+530}_{-80}$ & 3070$^{+990}_{-5}$ \\
\cii  & 11.1$^{+1.2}_{-1.3}$  & $\cdots$ & $\cdots$ \\
\civ & 472 $^{+3}_{-3}$  &    610$^{+25}_{-8}$ & 1160$^{+290}_{-7}$ \\
\nv  & 637$^{+4}_{-4}$  &    834$^{+9}_{-8}$ & 1360$^{+540}_{-5}$ \\
\ovi & 2700$^{+20}_{-20}$  &   3390$^{+1040}_{-20}$ &  4170$^{+1330}_{-4}$ \\
\siIII & 2.44$^{+0.06}_{-0.06}$ & $\cdots$ & $\cdots$ \\
\siiv & 11.9$^{+0.7}_{-0.6}$    &    14.1$^{+0.6}_{-0.6}$ & 15.1$^{+0.6}_{-0.6}$ \\
\siv & 101$^{+4}_{-4}$  & $\cdots$ & $\cdots$ \\

 \enddata
 \tablenotetext{~}{{\bf Notes.} Column densities are given as linear values in units of $10^{12}$ cm$^{-2}$. Asymmetrical errors are from photon statistics only and are derived using the technique outlined in \citet{Gabel05}.}

 \label{tabcold2}
\end{deluxetable}

\clearpage

\subsubsection{Effects of Deconvolution on Column Density Determinations}

Early on-orbit analysis of the COS Line Spread Function (LSF) revealed a significant departure
from the Gaussian profile observed during ground testing, showing
a broadened central core associated with strong broad wings scattering
a significant part of the flux (e.g. Kriss 2011 COS Instrument Science Report 2011-01(v1))\footnote{found at http://www.stsci.edu/hst/cos/documents/isrs/ISR2011\_01(v1).pdf}.
The major effect of such a degraded LSF is to scatter photons from the
continuum region into the absorption troughs, decreasing their
apparent depth and hence affecting the determined column density
derived for each ion, especially for narrower absorption troughs with $\Delta v$ $\sim$50~km~s$^{-1}$ (Kriss et al. 2011, accepted for publication in A\&A).

Given the high signal to noise of our spectrum, we attempted
to correct for the degraded LSF by deconvolving the spectrum using
the Lucy-Richardson algorithm implemented in the STSDAS routine 
\texttt{lucy} in the \texttt{analysis.restore} package. Given the wavelength dependance
of the LSF, the spectrum was deconvolved in 50~\AA\ intervals around the observed absorption troughs using the appropriate LSF for each region (see Kriss et al. 2011 for details). Column densities are then 
computed for the deconvolved absorption troughs after rebinning the data to the identical
20~km~s$^{-1}$ velocity grid. For the partial covering method, the typical increase in the computed 
column density for the broadest troughs (e.g. component 2 of \civ)
is of the order of several percent, but can reach up to $\sim$30\% for the narrower troughs such as \siiv.
While these results show a general increase in the measured column
densities on the deconvolved line profile, the significant increase
of the associated noise in the continuum implied by the deconvolution
process and the inability to accurately characterize that noise increase lead us
to consider the column densities derived on the non-deconvolved spectrum
in the remaining analysis while bearing in mind that the column densities associated
with narrow troughs can be underestimated by up to $\sim$~30\% for the partial covering solution due to the
degraded LSF.


\section{Photoionization Analysis}

\subsection{Spectral Energy Distribution}
\label{sec:sed}

The photoionization and thermal structure of an outflow depends on the spectral energy distribution (SED) incident upon the outflow. A standard SED (hereafter MF87) used in photoionization modeling of AGN was developed in \citet{Mathews87}. This SED contains a prominent ``big blue bump,'' which came under scrutiny in later works. For example, \citet{Laor97} dispute the existence of an X-ray-to-ultraviolet (XUV) bump in AGN based on ROSAT data. However, X-ray data of IRAS~F04250-5718 seem to require an XUV bump \citep{Turner99}. Since this object's SED is unobservable in the UV ionizing region, we analyze models with different XUV bumps. Studies of the SED in the X-ray regime have shown extreme variability (e.g. Guainazzi et al. 1998). Pounds et al. (2004a) showed that the variability could be explained by absorption of a relatively constant SED by cold dense matter. In order to ameliorate this issue, we use the \textit{simultaneous} observations from HST/COS and XMM-Newton to constrain the SED of IRAS F04250-5718: HST/COS data constrain the SED in the frequency range $2~\times~10^{15}~\textrm{Hz}$~$\ltsim$~$\nu$~$\ltsim$~$3~\times~10^{15}~\textrm{Hz}$, while data from XMM-Newton (Costantini et al. 2011, in preparation) constrain the high energy portion ($\nu \gtsim 10^{17}~\textrm{Hz}$) of the SED incident upon the UV absorber. While the FUSE data were taken four years earlier, they agree with the COS data in the overlap region. The co-added COS and FUSE data can be fit with a single power law (see section 3.1) extending the far UV constraints to $\sim$1~Ry.

\begin{figure}[ht]
 \caption{SEDs used in the analysis of IRAS~F04250-5718. Ionization potentials of \hi\ and \ovi\ are marked with dotted black lines to mark the UV ionizing region of the SEDs. The MF87 SED drops to a minimum of about $10^{43.5}$ erg s$^{-1}$ at E $\approx$ 365 eV.
 \vspace{5 mm}}
 \includegraphics[angle=90,width=1.0\textwidth]{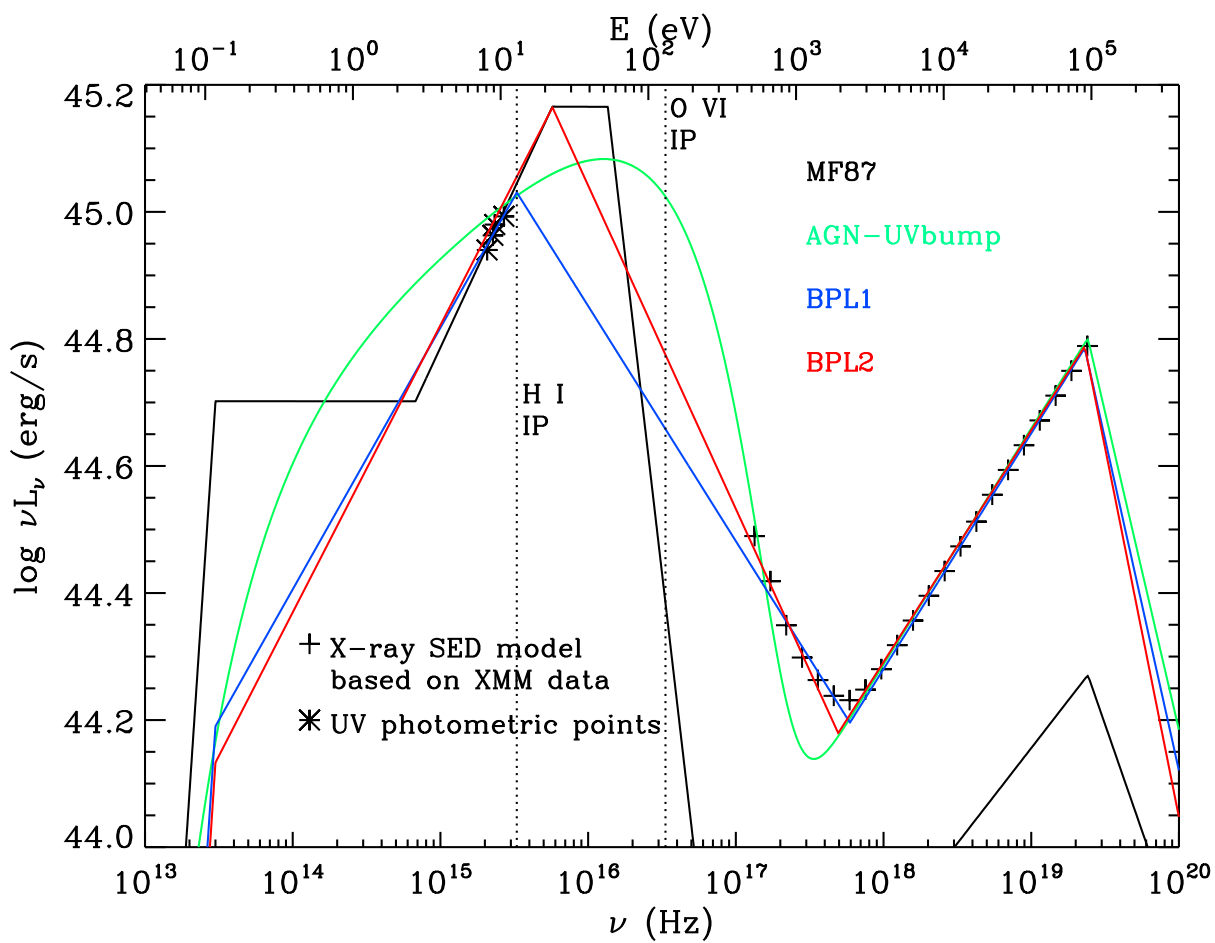}
 \label{sed_plot}
\end{figure}

We consider two families of SEDs in the analysis of IRAS~F04250-5718 for a total of three SEDs. Figure~\ref{sed_plot} shows the three SEDs along with the MF87 SED for comparison. For two of our SEDs, we use piecewise power law models. The blue line in Figure~\ref{sed_plot} (hereafter BPL1) is the softest SED we consider physically reasonable given the constraints imposed by the data. It has a spectral break at $\sim$1~Ry, very near the edge of detection in the far UV. The red line in Figure~\ref{sed_plot} (hereafter BPL2) represents the broken power law model that best fits the data while containing an blue-bump that is no more prominent than that in MF87.

For our final SED (hereafter AGN-UVbump), we generate the sum of two continua using the \texttt{agn} command in {\sc Cloudy}. This command produces the continuum given by
\begin{equation}
 f_\nu = a\nu^{\alpha_{uv}} \textrm{exp} \left(\frac{-h\nu}{kT_{BBB}}\right) + b\nu^{\alpha_x} \textrm{,}
\end{equation}
where $\nu$ is the frequency, $k$ is Boltzmann's constant, and $h$ is Planck's constant. The spectral index $\alpha_{uv}$ gives the low-energy slope of the optical/UV component, $\alpha_x$ gives the X-ray slope, and $T_{BBB}$ is the temperature of the ``big blue bump.'' The coefficient mutliplying the optical/UV power law component provides an exponential cutoff for energies $\ltsim 0.01$~Ry, and the coefficient multiplying the X-ray power law component is adjusted to produce the optical to X-ray slope ($\alpha_{ox}$) provided by the user. The X-ray component is set to zero for energies $\le$~0.1~Ry and falls off as $\nu^{-2}$ for energies $\gtsim$ 100 keV. Generated by the {\sc Cloudy} command \texttt{agn 2.8e6 -1.30 -0.80 -0.63} (where the numbers are $T_{BBB}$, $\alpha_{ox}$, $\alpha_{uv}$, and $\alpha_x$, respectively), the SED AGN-UVbump is similar to the one used by \citet{Arav07} for Markarian 279 with the main differences being the frequency at which the spectral break of the big blue bump occurs and the hard X-ray excess seen in the XMM-Newton data (also observed in \textit{Suzaku} data by \citealt{Turner09}).

Table \ref{lboltable} shows the bolometric luminosity ($L_{bol}$) and the number of hydrogen ionizing photons emitted by the source per second ($Q_H$) for each SED.

\begin{deluxetable}{lcccc}
 \tablewidth{0.8\textwidth}
 \tablecolumns{5}
 \tabletypesize{\footnotesize}
 \tablecaption{SED Emission Properties}
 \tablehead{
 \colhead{SED} &
 \colhead{$L_{bol}$ (10$^{45}$ erg/s)} &
 \colhead{$Q_H$ (10$^{55}$ s$^{-1}$)}
  }
 \startdata
 MF87 & 6.4 & 5.2 \\
 AGN-UVbump & 8.8 & 5.0 \\
 BPL1 & 6.6 & 3.6 \\
 BPL2 & 7.3 & 5.0 \\
 \enddata
 \label{lboltable}
\end{deluxetable}

\clearpage

\subsection{Photoionization Modeling}
The total hydrogen column density $N_H$ and ionization parameter
\begin{equation}
\label{eqionparam}
U_H \equiv {Q_H\over {4\pi R^2 \vy{n}{H} c}},
\end{equation}
for the absorber are determined by self-consistent photoionization modeling (where $c$ is the speed of light, $Q_H$ is the rate of hydrogen ionizing photons emitted by the AGN, $R$ is the distance to the absorber from the central source, and $n_H$ is the total hydrogen number density). For this, we use version c08.00 of the spectral synthesis code {\sc Cloudy}, last described by \citet{Ferland98}. In these models, we assume solar abundances as given in {\sc Cloudy} and a plane-parallel geometry for a gas of constant hydrogen number density. In $\S$\ref{sec:metallicity}, we discuss the effects of supersolar metallicities on our results. In determining the ionization parameter that leads to our distance determination, we assume that the high ionization features (\civ, \nv, \ovi) come from the same gas as the lower ionization species (\cii, \siIII, \siiv, \siv). This is supported by the kinematic correspondence for all of the ions in detected in the outflow (see Figure~\ref{figlineprof}).

\begin{figure}[ht]
 \caption{Component 1: Plots of constant $N_{ion}$ in the $N_H$-$U_H$ plane. The plotted lines trace models whose predicted $N_{ion}$ matches the observed values. The solid black square marks the model that best fits all of the given constraints. The horizontal dashed line at log $N_H = 24$ is the boundary above which an absorber becomes optically thick due to electron scattering. The slanted dashed line is the approximate location of the hydrogen ionization front. The error bars mark solutions that are within a factor of $\sim 2$ for all ions. $N_H$ and $U_H$ are correlated such that increasing $U_H$ requires an increase in $N_H$ with the line of solutions running roughly parallel to the \nv\ line from the center of the filled square.}
 \includegraphics[angle=90,width=1.0\textwidth]{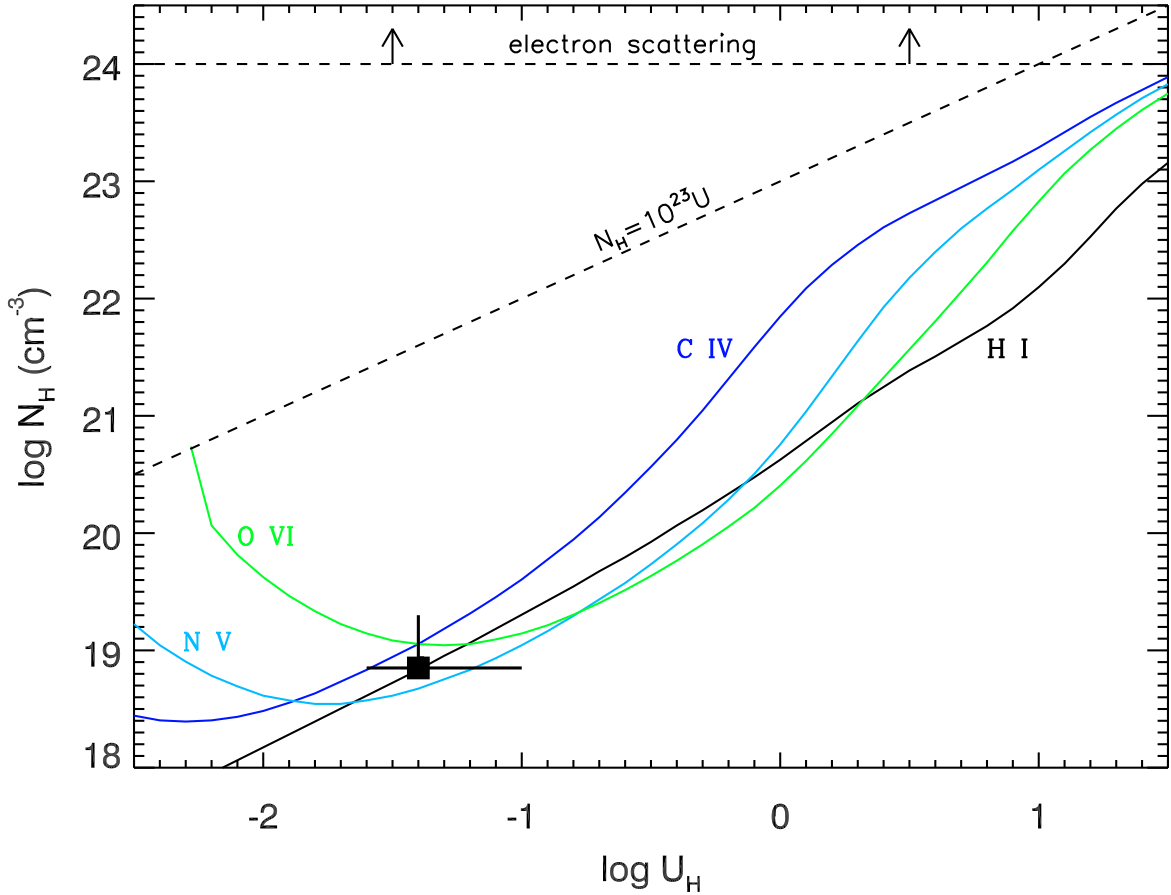}
 \label{low_v_grid}
\end{figure}
\begin{figure}[ht]
 \caption{Component 2+3: Plots of constant $N_{ion}$ in the $N_H$-$U_H$ plane. The plotted lines trace models whose predicted $N_{ion}$ matches the observed values. Symbols mark the models that fit certain constraints. The plus is equidistant from \cii, \civ, and \ovi. The square and diamond mark the crossing points of silicon ions and carbon ions, respectively. The best by-eye fit is marked with a triangle. The lower limit on \hi\ is due to the fact that Ly$\beta$ may be saturated, and the upper limit is determined from non-detection of a Lyman break (see $\S$\ref{sec:h1col}). The shaded band therefore represents values of \hi\ consistent with the data.}
 \includegraphics[angle=90,width=1.0\textwidth]{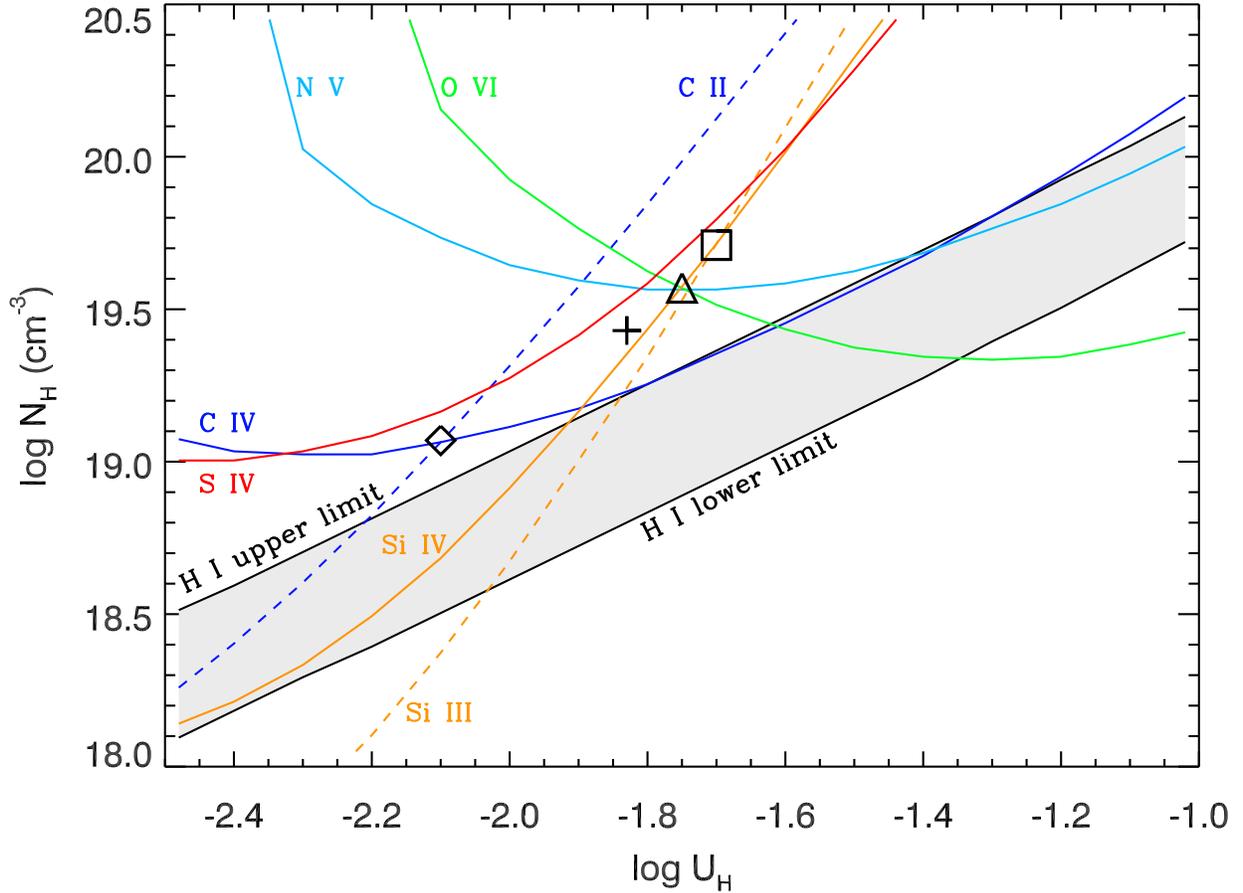}
 \label{high_v_grid}
\end{figure}

To determine $N_H$ and $U_H$, we generate grids of models varying these parameters in 0.1 dex steps (similar to the approach of \citealt{Arav01}). For a given SED, $\sim$4500 models are computed to investigate a parameter space with 15 $\le$ log $N_H$ $\le$ 24.5 and -5 $\le$ log $U_H$ $\le$ 2. At each point in a grid, we tabulate the predicted column densities of all relavent ions and then compare the measured column densities to those predicted by the models. Figures \ref{low_v_grid} and \ref{high_v_grid} show results of this analysis for kinematic components 1 and 2+3 of the outflow, respectively, using the SED AGN-UVbump developed in $\S$\ref{sec:sed}. In these figures, lines where a linear interpolation of the tabulated $N_{ion}$ matches the measured $N_{ion}$ are plotted in the $N_H$-$U_H$ plane.

In spectra with absorption features from three or more lines from the same energy level of the same ion, it is possible to test the different absorber models \citep{deKool02,Arav05}. However, in IRAS~F04250-5718, we have no observational constraints on the spatial distribution of absorbing material across the emission source. Therefore, we use the average value of $N_{ion}$ determined by partial covering and power law methods for \civ, \nv, ,\ovi, \hi, and \siiv. For the remaining ions (\cii, \siIII, and \siv), we have ionic column densities determined by the AOD method only. The fact that all three measurement methods give similar results for the column density of \siiv\ indicates the AOD method yields a reliable estimate of the actual column density for this ion. We assume this is the case for the other ions with shallower troughs since the line profiles are similar.

The first kinematic component has only four constraints since none of the low-ionization species are detected there. An approximate solution is determined by visual inspection of the graph and plotted as a solid square in Figure~\ref{low_v_grid} at log $U_H \sim$ -1.4 and log $N_H \sim$ 18.9. The error bars span the values of $N_H$ and $U_H$ that yield $N_{ion}$ to within a factor of $\sim 2$. The parameters are correlated such that increasing $U_H$ requires an increase in $N_H$ resulting in a line of solutions that runs roughly parallel to the curve plotted for \nv\ from the center of the solid square. Table~\ref{low_v_models} details the results of models using different SEDs for component 1 of the outflow.

 
Figure \ref{high_v_grid} shows the results for kinematic component 2+3 using the SED AGN-UVbump. Four different solutions are given: The solution marked with a plus is approximately equidistant from the \cii, \civ, and \ovi\ curves and encompasses all lines to within a factor of $\sim 2.5$. The crossing point of \siIII\ and \siiv\ is marked with a square, and the crossing point of \cii\ and \civ\ is marked with a diamond. Finally, our best by-eye fit to the data is marked with a triangle at log $U_H \sim -1.8$ and log $N_H \sim 19.6$. The ``equidistant fit'' and the ``best by-eye fit'' both provide reasonably good fits to the data. We prefer the best by-eye fit since it obeys the firm lower limit determined by $N$(\ovi) while fitting all of the ions except \cii\ to within a factor of $\sim 2.2$ as well as improving the fits to \nv\ and \siiv, while giving results for \civ, \siIII, \siv, and \hi\ that are similar to those given for the equidistant fit. The disadvantage of the best by-eye fit is that $N$(\cii) is underpredicted by a factor of $\sim 5$ while for the equidistant fit, $N$(\cii) is underpredicted by a factor of $\sim 2.2$. Table~\ref{high_v_models} details the results of models using different SEDs for component 2+3 of the outflow.

Solutions for all three SEDs are compared in Figure~\ref{fig:solutions}. A dashed box encloses the solution space that gives reasonable fits to the data while allowing for changes in SED shape and chemical abundances. Solutions determined by the crossing of \cii\ and \civ\ in the $N_H$-$U_H$ plane are omitted because they severely underpredict \ovi, for which we have a firm lower limit. Less severe underpredictions may be alleviated by supersolar metallicities, which we investigate in the next section.

\begin{deluxetable}{lcccc}
 \tablewidth{0.8\textwidth}
 \tablecolumns{5}
 \tabletypesize{\footnotesize}
 \tablecaption{Photoionization models for component 1 (-90~km~s$^{-1}$ $\le $ v $\le 30$~km~s$^{-1}$)}
 \tablehead{
 \colhead{Ion} &
 \colhead{log($N_{ion}$) (cm$^{-2}$)} &
 \multicolumn{3}{c}{log$\left(\frac{N_{mod}}{N_{obs}}\right)$} \\
 \colhead{} &
 \colhead{observed $^\dagger$} &
 \colhead{AGN-UVbump} &
 \colhead{BPL1} &
 \colhead{BPL2} 
  }
 \startdata
 log $U_H$ & $\cdots$ & -1.40 & -1.23 & -1.21 \\
 log $N_H$ & $\cdots$ & 18.85 & 18.90 & 18.87 \\
\hline
 H I & 14.97$^{+0.16}_{-0.24}$ & +0.02 & -0.21 & -0.19 \\
 C IV & 14.32$^{+0.16}_{-0.27}$ & -0.21 & -0.21 & -0.21 \\
 N V & 14.02$^{+0.11}_{-0.16}$ & +0.18 & +0.20 & +0.20 \\
 O VI & 15.29$^{+0.13}_{-0.20}$ & -0.20 & -0.17 & -0.19 \\

 \enddata
 \tablenotetext{\dagger}{Average of column densities derived from the partial covering and power law methods. Uncertainties reflect the range of column densities with the higher given by the power law method and the lower given by the partial covering method.}
 \label{low_v_models}
\end{deluxetable}

\begin{deluxetable}{lcccc}
 \tablewidth{0.8\textwidth}
 \tablecolumns{5}
 \tabletypesize{\footnotesize}
 \tablecaption{Photoionization models for component 2+3 (-290~km~s$^{-1} \le$ v $\le -90$~km~s$^{-1}$)}
 \tablehead{
 \colhead{Ion} &
 \colhead{log($N_{ion}$) $^\ddagger$ (cm$^{-2}$)} &
 \multicolumn{3}{c}{log$\left(\frac{N_{mod}}{N_{obs}}\right)$} \\
 \colhead{} &
 \colhead{observed} &
 \colhead{AGN-UVbump} &
 \colhead{BPL1} &
 \colhead{BPL2} 
  }
 \startdata
 log $U_H$ & $\cdots$ & -1.75 & -1.60 & -1.57 \\
 log $N_H$ & $\cdots$ & 19.58 & 19.59 & 19.57 \\
\hline
 H I $^*$ & 15.41-15.82 & +0.31 & +0.06  & +0.08 \\
 C II $^\ddagger$ & 13.05$^{+0.04}_{-0.05}$ & -0.42 & -0.50  & -0.67 \\
 C IV $^\dagger$ & 14.95$^{+0.11}_{-0.16}$ &     +0.29  & +0.29 & +0.28  \\
 N V $^\dagger$ & 15.04$^{+0.09}_{-0.12}$ &      +0.02  & +0.01 & +0.01  \\
 O VI $^\dagger$ & 15.58$^{+0.04}_{-0.05}$ &      +0.01 & -0.01 & -0.03  \\
 Si III $^\ddagger$ & 12.39$^{+0.01}_{-0.01}$ & +0.03 & +0.04 & +0.03 \\
 Si IV $^\dagger$ & 13.16$^{+0.02}_{-0.01}$ &      +0.05 & -0.03 & +0.06 \\
 S IV $^\ddagger$ & 14.00$^{+0.02}_{-0.01}$  & -0.10 & -0.14 & -0.11 \\
\enddata
 \tablenotetext{*}{Upper limit determined from the non-detection of a Lyman break in FUSE data. The range of values for \hi\ is shown as a gray strip in Figure~\ref{high_v_grid}.}
 \tablenotetext{\ddagger}{Column densities derived from the AOD method. Uncertainties are due to photon statistics only and are thereby underestimated.}
 \tablenotetext{\dagger}{Average of column densities derived from the partial covering and power law methods. Uncertainties reflect the range of column densities with the higher given by the power law method and the lower given by the partial covering method.}
 \label{high_v_models}
\end{deluxetable}

\clearpage

\subsubsection{Supersolar Metallicity Models}
\label{sec:metallicity}
\begin{figure}[ht]
\caption{Photoionization solutions for component 2+3 of the outflow for different SEDs and metallicities. The colors match those of the SEDs in Figure~\ref{sed_plot}: Green is the AGN-UVbump SED, blue is BPL1, and red is BPL2. Abundances are solar unless noted otherwise: Solar abundances are from {\sc Cloudy}, and supersolar abundances are from \citet{Ballero08}. Only the best by-eye fits are shown for supersolar metalllicites. The dashed box shows the range of plausible solutions.}
  \includegraphics[angle=90,width=1.0\textwidth]{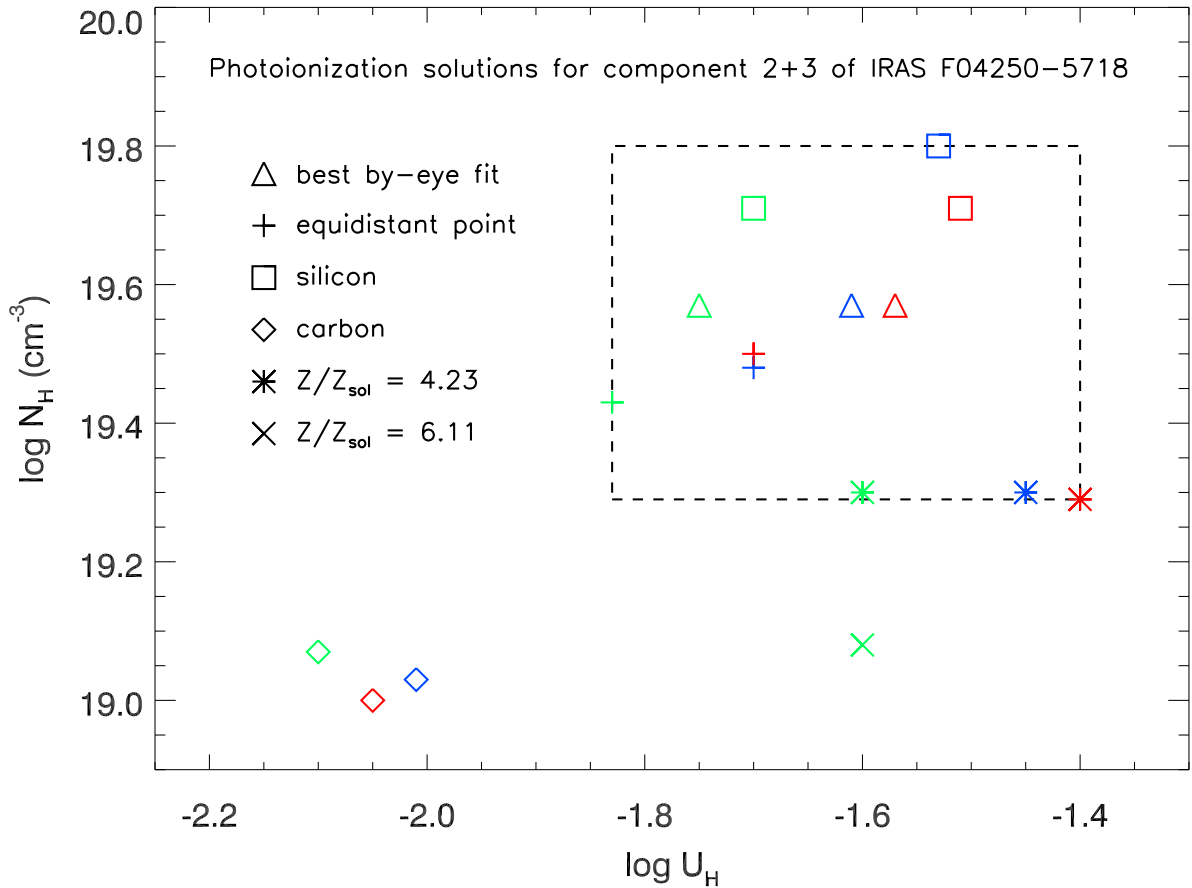} \\
 \label{fig:solutions}
\end{figure}

The ionization parameters for the outflow were determined under the assumption of solar abundances. The grids of models support this assumption for kinematic component 1 of the outflow since the metal lines have crossing points very close to the \hi\ line, but kinematic compononent 2+3 shows signs of supersolar metallicities. In particular, the equidistant solution and the best by-eye fit solution both overpredict \hi: In Figure~\ref{high_v_grid}, these solutions lay above the line marking the \hi\ upper limit. In order to investigate the effects of changes in metallicity ($Z$) on our determination of $N_H$ and $U_H$, we use the abundances given by \cite{Ballero08} for $Z/Z_\odot$=4.23 and 6.11. They also list abundances for $Z/Z_\odot$=7.22, but these do not yield a good solution for any of our SEDs. The results are plotted in Figure~\ref{fig:solutions}. For each SED, there is an increase in $U_H$ of $\sim$0.1 dex and a decrease in $N_H$ of up to $\sim$0.4 dex. We note that the solutions for $Z/Z_\odot$=4.23 were generally better than those for $Z/Z_\odot$=6.11, which underpredicted \hi\ for both of the broken power law SEDs. For the AGN-UVbump SED, the solution found using $Z/Z_\odot$=6.11 is plotted but left out of the box of reasonable solutions since the fit was so poor. The \civ\ was fit better in the supersolar models than in the solar ones due to a relative increase in abundance of silicon and nitrogen with respect to carbon, but this resulted in an underprediction of \cii\ that is more severe than in the solar models. In fact, for the AGN-UVbump SED, using $Z/Z_\odot$=4.23, all lines are fit to within a factor of $\sim 1.1$ except for \cii\ whose column density is underpredicted by a factor of $\sim 10$. We defer the full determination of chemical abundances for the outflow in IRAS~F04250-5718 to a future paper.

\clearpage

\subsubsection{Multiple Ionization Parameter Models}

For optically thin models, multiple ions from the same chemical species offer (approximately) metallicity independent solutions for $N_H$ and $U_H$. In the COS spectrum of IRAS~F04250-5718, we have \cii, \civ, \siIII, and \siiv. The results obtained using just the ratios of $N$(\cii)/$N$(\civ) and $N$(\siIII)/$N$(\siiv) are shown in Figure~\ref{fig:solutions}: The diamonds mark the results from carbon and the squares mark the results from silicon. While the values of $N_H$ agree very well, there is a discrepancy in the determination of $U_H$ of $\sim 0.4$~dex. Furthermore, the ratio $N$(\civ)/$N$(\cii) is substantially overpredicted by the models that give reasonable fits to the other ions. We therefore investigate the possibility of two ionization parameters improving on the photoionization solutions for kinematic component 2+3.

For all three SEDs, the by-eye best fits (marked with triangles in Figure~\ref{fig:solutions}) underpredict the column densities of \cii\ and \siIII. By choosing a second ionization parameter that is near the crossing point (log $U_H \sim -3.5$) of \cii\ and \siIII\ in the $N_H$-$U_H$ plane, we improve the fits to these two low ionization ions while leaving essentially unchanged the predicted column densities of the other ions when values from the two models are summed. However, low ionization parameter models predict a substantial amount of \siII, which we do not detect in the data. Using an upper limit of log $N$(\siII) $\sim 11.9$~cm$^{-2}$, the two ionization parameter solutions overpredict $N$(\siII) by $\sim 0.6$ dex. We conclude that the two ionization parameter does not significantly improve our results, simply trading the underprediction of $N$(\cii) for the overprediction of $N$(\siII).


\section{Absorber Distance and Energetics}

\begin{figure}
 \includegraphics[angle=90,width=1.0\textwidth]{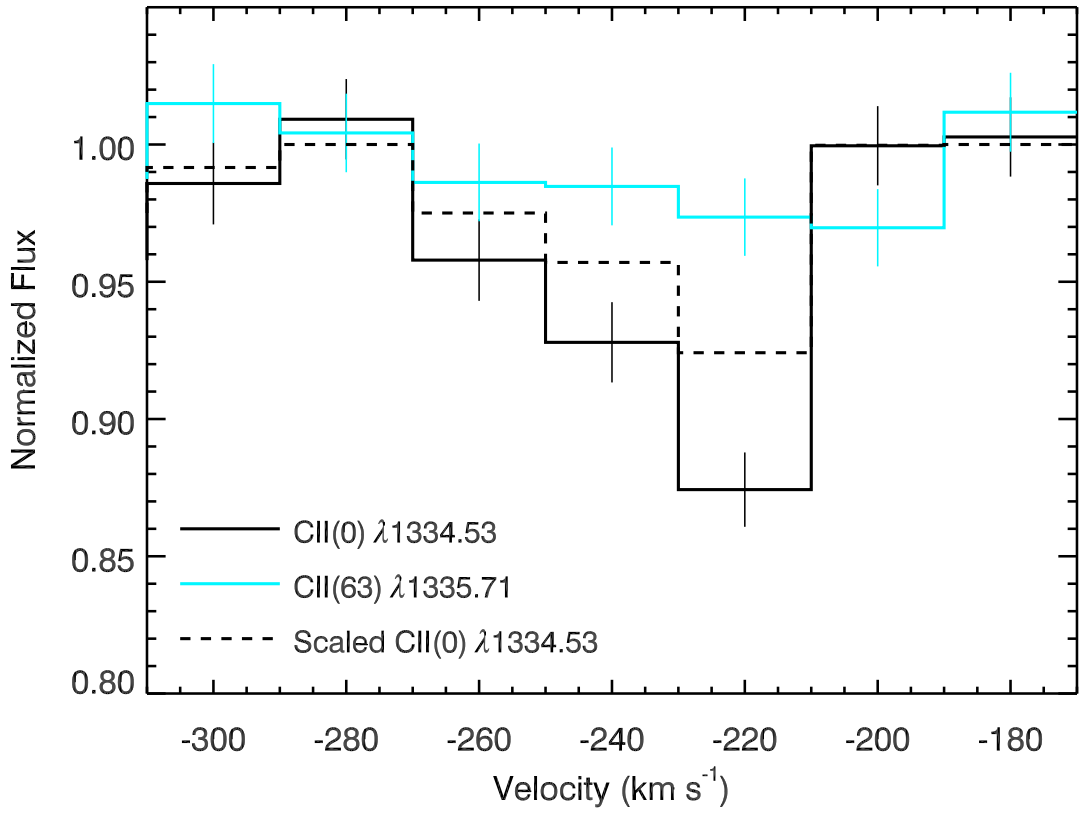}
 \caption{Line profiles for \cii\ and \cii* rebinned to 20 km/s. We place an upper limit on \cii* by assuming the noise could hide up to a 3-$\sigma$ detection. The dashed line shows the \cii\ trough scaled to the 3-$\sigma$ noise in the region where \cii* would be.\label{fig:ciiprofile}}
\end{figure}

\begin{figure}
 \includegraphics[angle=90,width=1.0\textwidth]{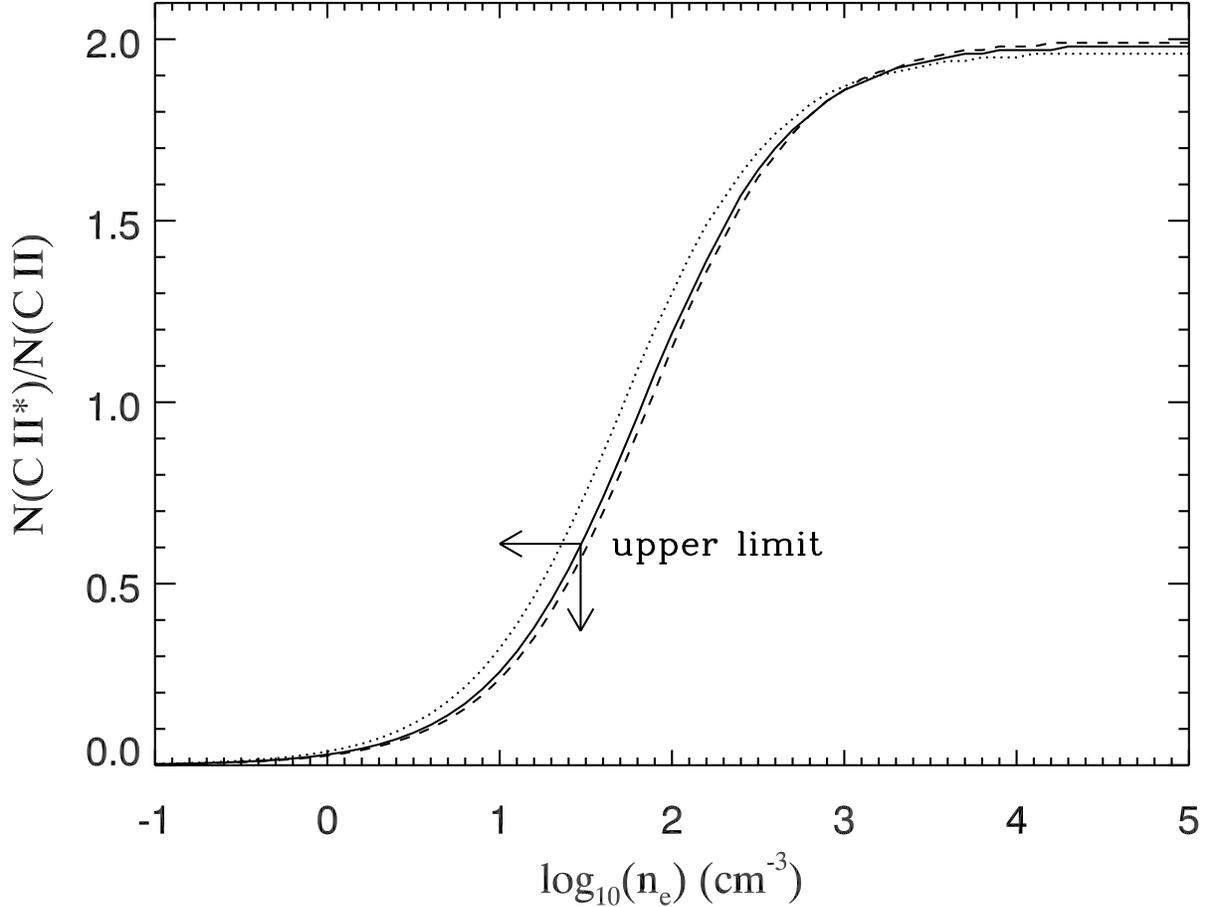}
\caption{The ratio of the level population of the first excited state ($E=63$ cm$^{-1}$) to the level population of the ground state for \cii\ versus electron number density ($n_e$) for T = 10000 K (dotted line), T = 15000 K (solid line), and T = 30000 K (dashed line). Non-detection of \cii* implies a level population ratio of $\sim 0.25 \pm 0.12$. For a conservative upper limit, we use the 3-$\sigma$ value of 0.61 yielding $n_e < 30$~cm$^{-3}$. \label{fig:ciidens}}
\end{figure}

The distance to the absorber can be computed from the ionization parameter (see Equation~(\ref{eqionparam})) if the total hydrogen number density $\vy{n}{H}$ is known. To estimate $\vy{n}{H}$ in kinematic component 3 (the high velocity component) we consider the ratio of populations of the first excited metastable and ground levels of \cii. The rebinned line profiles of \cii\ $\lambda1335$ and \cii* $\lambda1336$ are shown in Figure \ref{fig:ciiprofile}. 
Integrating the flux in the region where a \cii* trough should be according to the detected \cii\ trough yields a column density for \cii* of $2.81^{+1.31}_{-1.07} \times 10^{12}$~cm$^{-2}$ where the errors are 1-$\sigma$ photon statistics. Due to non-detection of the excited state of \cii, we place a conservative upper limit on the column density of \cii* of $6.74 \times 10^{12}$~cm$^{-2}$ by assuming the noise could hide up to a 3-$\sigma$ detection. 
Comparing collisional excitation models with the ratio of the column density measurement of \cii\ ($11.1 \pm 1.3 \times 10^{12}$ cm$^{-2}$) and the upper limit on the column density of \cii* ($\ltsim 6.74 \times 10^{12}$ cm$^{-2}$), we place an upper limit on the free electron number density $n_e \ltsim$ 30 cm$^{-3}$ (see Figure~\ref{fig:ciidens}). Since our photoionization modeling implies we are within the \heiii\ region, $n_e$ $\approx$ 1.2$\vy{n}{H}$. 
In order to place robust lower limits on the distance, mass flow rate, and kinetic luminosity of the absorber, we adopt the model with the SED BPL2 and a metallicity of $Z = 4.23$ which yields the maximum $U_H$ ($10^{-1.4}$) and minimum $N_H$ ($10^{19.3}$) of all our models. 
These results imply a distance of
\begin{equation}
 R = \left (\frac{Q_H}{4\pi c U \vy{n}{H}} \right)^{1/2} \gtsim 3 \textrm{ kpc}
\end{equation}
for kinematic component 2+3. We note that the data hint at a detection of \cii*. Since the \cii* trough that appears in Figure~\ref{fig:ciiprofile} is too shallow to clearly distinguish from the noise, we do not claim we have a measurement of the column density of \cii*, but state that the lower limit on the distance may be an estimation of the actual distance. We cannot determine the distance to kinematic component 1, since we have no density diagnostics for this component.

In order to estimate the mass, mass flow rate, and kinetic luminosity of the outflow, we must assume a geometry. For a full discussion of the geometry used here, see \citet{Arav11}. Assuming that the outflow has the geometry of a thin, partially-filled shell, the mass is given by
\begin{equation}
 M_{out} = 4\pi \mu m_p R^2 N_H \Omega \gtsim 2\times10^7 \Omega_{0.5} M_{\odot},
\end{equation}
where $\mu = 1.4$ is the mean atomic mass per proton, $m_p$ is the mass of the proton, and $\Omega$ is the covering fraction of the outflow as seen from the central source. We adopt the value $\Omega = 0.5$ due to the fact that outflows are seen in about 50\% of observed Seyfert galaxies \citep{Dunn07} and the luminosity of IRAS~F04250-5718 is just above that of a Seyfert galaxy. A comprehensive discussion of covering fraction for quasar outflows is given in \citet{Dunn10a}. In order to reflect the uncertainty in $\Omega$, we provide our results in units of $\Omega_{0.5} = \Omega/0.5$. The average mass flow rate is just the total mass divided by the dynamical timescale $R/v$:
\begin{equation}
 \dot{M}_{out} = 4\pi \mu m_p R N_H v \Omega \gtsim 1 \textrm{ } \Omega_{0.5} \textrm{ } M_{\odot} \textrm{yr}^{-1}.
\end{equation}
This value is similar to the mass accretion rate we find under the assumption that the black hole is accreting mass at the Eddington rate \citep{Salpeter64}:
\begin{equation}
 \dot{M}_{acc} = \frac{L_{bol}}{\epsilon c^2} \sim 2 \textrm{ } M_{\odot} \textrm{yr}^{-1},
\end{equation}
where we have assumed an efficiency $\epsilon$ = 0.1.
Finally, the kinetic luminosity is given by
\begin{equation}
 \dot{E}_k = \frac{\dot{M}_{out} v^2}{2} \gtsim 1 \times 10^{40} \textrm{ } \Omega_{0.5} \textrm{ } \textrm{erg~s}^{-1}.
\end{equation}
In Table \ref{table:objects}, we compare these results with those we have found in high luminosity quasars.

Given that the bolometric luminosity is $L_{bol} \sim 9 \times 10^{45}$ erg s$^{-1}$, this outflow is not energetically significant in AGN feedback scenarios, which typically require $\dot{E}_k$ to be a few percent of $L_{bol}$ (e.g., \citealt{Scannapieco04}). The mass flow rate is also rather small for the purpose of chemically enriching the environment. Assuming a duty cycle of $10^8$ for the active phase of this AGN, Hellman and Arav (2011, in preparation) find that a mass flow rate of 1 solar mass per year is only enough to enrich a small galactic halo of $\sim 10^{10} M_\odot$ to the observed Intra-Cluster values, or alternatively 1 Mpc$^3$ of the Inter-Galactic Medium. The outflows observed in the high luminosity objects shown in Table \ref{table:objects} are much more important for AGN feedback, both energetically and for chemical enrichment.

\begin{deluxetable}{lcccccccccc}
\rotate
 \tablewidth{1.0\textwidth}
\setlength{\tabcolsep}{0.05in}
 \tablecolumns{11}
\tabletypesize{\scriptsize}
 \tablecaption{Properties of Large Scale Outflows Measured by Our Group to Date\label{table:objects}}
 \tablehead{
 \colhead{Object} &
 \colhead{log $L_{bol}$} &
\colhead{log $n_e$} &
\colhead{$\overline{v}$ $^\dagger$} &
\colhead{$\Delta v$ $^\ddagger$} &
 \colhead{$R$} &
 \colhead{log $N_H$} &
 \colhead{log $U_H$} &
 \colhead{log $\dot{E}_k$} &
 \colhead{$\dot{M}_{out}$} &
 \colhead{References $^*$} \\
 \colhead{} &
 \colhead{(erg s$^{-1}$)} &
\colhead{(cm$^{-3}$)} &
\colhead{(km s$^{-1}$)} &
\colhead{(km s$^{-1}$)} &
 \colhead{(kpc)} &
 \colhead{(cm$^{-2}$)} &
 \colhead{} &
 \colhead{(erg s$^{-1}$)} &
 \colhead{($M_\odot$ yr$^{-1}$)} &
 \colhead{}
  }
 \startdata

IRAS~F04250-5718 & 45.9 & $< 1.4$ & -200 & 200 & $> 3.7$ & $> 19.3$ & $< -1.4$ & $> 40.13$ & $> 1.07$ & 1 \\
QSO 1044+3656 & 46.8 & 3.8 & -3995 & 420 & 1.7 $\pm$ 0.4 & 20.8 $\pm$ 0.1 & -2.19 $\pm$ 0.10 & 44.81$^{+0.09}_{-0.11}$ & 120 $\pm$ 25 & 2 \\
AKARI J1757+5907 & 47.6 & $< 3.8$ & -925 & 250 & $> 3.7$ & $> 20.8$ & $>-2.2$ & $> 43.30$ & $> 70$ & 3 \\
SDSS J0318-0600 & 47.7 & 3.3 & -4160 & 890 & 5.9 $\pm$ 0.4 & 19.9 $\pm$ 0.2 & -3.1 $\pm$ 0.1 & 44.55$^{+0.10}_{-0.15}$& 60 $\pm$ 20 & 4 \\
SDSS J0838+2955 & 47.5 & 3.8 & 5000 & 2000 & 3.3$^{+1.5}_{-1.0}$ & 20.8 $\pm$ 0.3 & -2.0 $\pm$ 0.2 & 45.35$^{+0.23}_{-0.22}$ & 300$^{+210}_{-120}$ & 5 \\
QSO 2359-1241 & 47.7 & 4.4 & 1385 & 130 & 3.2$^{+1.8}_{-1.1}$ & 20.6 $\pm$ 0.2 & -2.4 $\pm$ 0.2 & 43.36 $\pm$ 0.27 & 90$^{+35}_{-20}$ & 6 \\

 \enddata
\tablenotetext{\dagger}{Mean velocity of the kinematic components used in the analysis.}
\tablenotetext{\ddagger}{Full-width at zero-intensity.}
 \tablenotetext{*}{(1)~This work, (2)~\citet{Arav11}, (3)~\citet{Aoki11}, (4)~\citet{Dunn10a}, (5)~\citet{Moe09}, (6)~\citet{Arav08,Korista08}.}
 \label{table:objects}
\end{deluxetable}

\section{Discussion}

\subsection{Comparison with Galactic Winds}

While it is clear that outflows at distances of $\ltsim$ 10 pc and/or with velocities $\gtsim 500$~km~s$^{-1}$ \citep{Martin05} are AGN driven, outflows at distances~$\gtsim$~1~kpc with lower velocities, such as the one we find in IRAS~F04250-5718, are difficult to distinguish from galactic winds. Galactic winds are ubiquitous in starburst galaxies: Studies show $\gtsim$ 75\% of ultra-luminous infrared galaxies (ULIRGs) have winds (\citealt{Veilleux05} and references therein). An optical study of edge-on Seyfert galaxies by \citet{Colbert96a} revealed large-scale outflows in $\sim$25\% of these AGN. However, since evidence of starburst activity is found in about half of optically selected Seyfert 2 galaxies \citep{Gonzalez-Delgado00}, it is difficult to determine whether the galactic-scale outflows in Seyferts are driven by the AGN or starbursts. In a sample of ultra-luminous infra-red galaxies (ULIRGs) containing both an AGN and a high star formation rate, \citet{Rupke05} found that the velocity distributions in outflows observed in Seyfert 2 galaxies differed from those found in starburst galaxies with no observable AGN activity, suggesting the AGNs play a role in driving the galactic-scale outflows in these objects.

Much of the work on galactic winds has focused on optical emission lines produced by the warm gas (e.g., \citealt{Colbert96a,Lehnert96}), and X-ray emission lines produced by the hot gas (e.g., \citealt{Dahlem98}). Emission line techniques favor edge-on galaxies where it is easier to separate the wind emission from background emission. Complementary studies of absorption lines (e.g., \citealt{Heckman00,Martin05,Martin09}) favor face-on galaxies. Absorption line studies of galactic winds afford the advantage that the whole range of gas densities in outflows can be probed unlike emission line studies where the densest gases are favored \citep{Heckman00}. Much of this work has relied on \nai\ $\lambda \lambda$5892,5898 lines and find they do not completely cover the source. \citet{Martin09} found they could better constrain the kinematics of the outflow using \mgii\ $\lambda \lambda$2796,2803 since the absorption features are well separated allowing for better analysis than the heavily blended \nai\ doublet.

The distance and speed we find for kinematic component 2+3 is typical for a galactic wind \citep{Veilleux05}. We are unable to determine whether this outflow is AGN or starburst driven, but partial-covering of the emission sources implies the outflow is intrinsic to the host galaxy so photoionization of the outflow is likely dominated by AGN emission. In this case, regardless of outflow's origin, photoionization analysis, along with the upper limit on the electron number density provided by a detection of \cii\ with no detection of \cii*, yields a lower limit on distance to the absorber from the central source.

\section{Summary}

The HST/COS spectrum of the low luminosity quasar IRAS~F04250-5718 reveals an outflow with both high and low ionization species. The lines are narrow (FWHM $\ltsim 200$ km s$^{-1}$), and similar redshifts for corresponding emission and absorption features implies the outlflow is intrinsic to the AGN. Further evidence of the intrinsic nature of this outflow is partial covering of the emission source evident in \civ, \nv, \ovi, and \hi.

Using the measured column density of \cii\ along with an upper limit on the column density of \cii*, we found a conservative upper limit on the electron number density in the outflow of $\sim 30$ cm$^{-3}$. Photoionization modeling yielded a range in ionization parameter log~$U_H \sim -1.8 \pm 0.2$ and total hydrogen column density log~$N_H \sim 19.55 \pm 0.25$ cm$^{-2}$. Using model assumptions that minimize the distance to the outflow, we determined a conservative lower limit of $R \gtsim 3$ kpc. Assuming a thin-shell geometry, lower limits on the mass flow rate and kinetic luminosity were found to be $\gtsim 1$ $M_{\odot}$yr$^{-1}$ and $\gtsim 1 \times 10^{40} \textrm{ } \textrm{erg~s}^{-1}$, respectively.

In determining limits on the distance, mass flow rate, and kinetic luminosity of the outflow, we investigated the effects on our results of changing the SED incident upon the absorber and the metallicity of the gas as well as differences in column density determinations when different absorber models (i.e. partial-covering versus power-law models) were assumed. The ionization parameters found for the three different SEDs are within $\sim 0.2$ dex of each other, while increasing the metallicity resulted in ionization parameters $\sim 0.2$ dex higher for each SED and lowered $N_H$ by $\sim 0.3$~dex from the best-fit solar models. Changing the SED had little effect on the total hydrogen column density, while differences in the absorber model had little effect on ionization parameter and changed $N_H$ by $< 0.1$~dex.

We acknowledge support from NASA STScI grants GO 11686  and GO 12022 as well as NSF grant AST 0837880.  We would also like to thank Pat Hall for insightful suggestions and discussions. JIGS and CB acknowledge financial support from the Spanish Ministerio de Ciencia e Innovacion under project AYA2008-06311-C02-02.

when changes in SED shape and chemical abundances are considered as well as differences in column density determinations when different absorber models (i.e. partial-covering versus power-law models) are used


\bibliographystyle{apj}

\bibliography{astro}{}


\end{document}